\documentclass[aps,prl,twocolumn,preprintnumbers,superscriptaddress,amsmath,amssymb,footinbib]{revtex4-2}
\usepackage{graphicx,epsfig}
\usepackage{bm}
\usepackage{dcolumn}
\usepackage[english]{babel}
\usepackage[utf8]{inputenc}
\usepackage{comment}
\usepackage{amsmath}
\usepackage{amsfonts}
\usepackage{amssymb}
\usepackage{color}
\usepackage[usenames,dvipsnames]{xcolor}
\usepackage[colorlinks,bookmarks=false,citecolor=NavyBlue,linkcolor=Red,urlcolor=blue,
]{hyperref}
\usepackage[export]{adjustbox}
\usepackage{amsthm}
\usepackage{simplewick}
\usepackage{bbold}
\usepackage{ wasysym }
\usepackage[title]{appendix}
\usepackage{mathrsfs}  
\usepackage{braket}
\usepackage{graphics}
\usepackage{todonotes}

\date{ \today} 

\DeclareUnicodeCharacter{2212}{-}

\begin{document}


\title{Matrix Product States with Backflow correlations}
\author{Guglielmo Lami}
\affiliation{International School for Advanced Studies (SISSA), 34136 Trieste, Italy}
\author{Giuseppe Carleo}
\affiliation{Ecole Polytechnique Fédérale de Lausanne (EPFL), Institute of Physics, CH-1015 Lausanne, Switzerland}
\author{Mario Collura$^{1}$}

\begin{abstract}
By taking inspiration from the backflow transformation for correlated systems, we introduce a novel tensor network ansatz which extend the well-established Matrix Product State representation of a quantum-many body wave function. This new structure provides enough resources to ensure that states in dimension larger or equal than one obey an area law for entanglement. It can be efficiently manipulated to address the ground-state search problem by means of an optimization scheme which mixes tensor-network and variational Monte-Carlo algorithms. We benchmark the new ansatz against spin models both in one and two dimensions, demonstrating high accuracy and precision. We finally employ our approach to study the challenging $S=1/2$ two dimensional $J_1 - J_2$ model, demonstrating that it is competitive with the state of the art methods in 2D. 
\end{abstract}
\maketitle


Understanding Quantum Many-Body (QMB) systems in and out of equilibrium is one of the most exciting open challenges in physics and chemistry. In recent years, significant progress has been made in the study of strongly correlated quantum systems, on many fronts. For example, several experimental approaches implementing Feynmans' quantum simulators \cite{feynman1982simulating} are allowing the controlled exploration of uncharted territory ~\cite{nature11596,nature18274,nature24622,Schauss_2018,science.aax9743, RevModPhys.80.885, Schafer2020}. 

On the theoretical level, the development of Tensor-Networks (TN) techniques has significantly expanded the scope of variational approaches to QMB systems since the introduction of the Density-Matrix Renormalization Group (DMRG) algorithm \cite{PhysRevLett.69.2863}. The goal of TNs is to represent the QMB wave functions by means of a set of local tensors, connected in a generic network via auxiliary bonds with finite dimension $\chi$, thus overcoming computational limitations due to the exponentially large Hilbert space \cite{silvi2019, 2020Tirrito}. The bond dimension $\chi$ can be adjusted to manipulate the information content of the TN, thus going from product states ($\chi \! = \! 1$), reproducing mean-field approximations, to the exact but inefficient wave function representation.
In 1D, the Matrix Product State (MPS) geometry has demonstrated an unprecedented degree of accuracy for both equilibrium and out-of-equilibrium problems \cite{schollwoeck2011,paeckel2019}. 
However, TN have some fundamental limitations, such as the intrinsic hardness of finding efficient contraction schemes \cite{Eisert2014} and unfavorable scaling of the required resources with the system size in higher dimension \cite{silvi2019}.
Most successful TN geometries, 
like Projected-Entangled Pair States (PEPS) \cite{Verstraete2004RenormalizationAF}
and Tree Tensor Networks (TTN) \cite{PhysRevA.74.022320}, suffer from specific drawbacks: while the latter does not satisfy the Entanglement area law (although some effort has been spent to overcome this limitation in Ref.~\cite{PhysRevLett.126.170603}), the former suffers from high algorithmic complexity, $O(\chi^{10})$, and lacks exact computation of expectation values. 

In parallel to the progress of TN, artificial Neural Networks (NN) have been discovered and used in a plethora of different scientific fields, proving astonishing versatility in physics applications~\cite{carleocirac2019}. In recent years, they have been employed as a variational ansatz for QMB problems~\cite{carleonature}. In this context, a number of possible architectures have been tried, such as Restricted Boltzmann Machine (RBM) \cite{carleonature,melko_restricted_2019}, Feed-Forward NN (FFNN) \cite{sharir2021neural, carleonaturecommunications2018} and Recurrent NN (RNN) \cite{PhysRevResearch.2.023358}. These ansatze have been proven to have a great descriptive power~\cite{sharir2021neural, Collura2021}. However, the number of parameters entering a NN wave function may be arbitrarily large and the appropriate network structure is usually not clear a priori. Understanding an optimal geometry encoding information from the specific dimensionality of the problem and taking advantage from both TN and NN structures could be the ultimate solution to the QMB problem. 

NN are usually optimized by means of variational Monte-Carlo (VMC) methods. Furthermore, a key tool in NN optimization is the so-called Automatic Differentiation \cite{BARTHOLOMEWBIGGS2000171}, which allows to efficiently compute cost-function derivatives with machine precision. This paradigm have been recently applied also to the TNs optimization \cite{PhysRevX.9.031041}. Combining such approaches with standard TN algorithms appears as a promising way to find new optimal strategies to solve open problems at the equilibrium and out-of-equilibrium. Efforts in this direction were made with the introduction of the Entangled Plaquette States (EPS)~\cite{Mezzacapo_2009,Mezzacapo_2010}, Monte-Carlo optimized PEPS \cite{LiuPhysRevB.95.195154} and infinite PEPS optimized with automatic differentiation \cite{10.21468/SciPostPhys.10.1.012}. The space of possible hybrid wave functions is however still largely unexplored.  

Here, we introduce a novel variational ansatz, generalizing the usual MPS. The ansatz is inspired by the so-called backflow techinque, commonly employed in electronic-structure theory \cite{PhysRev.102.1189, PhysRevB.78.041101, luo2019}. These new Matrix Product Backflow States (MPBS) can overcome some limitations of MPS by encoding an extensive amount of entanglement and keeping the algorithmic complexity under control. We further introduce a simple optimization scheme mixing DMRG and VMC recipes which can be proficiently applied to MPBS in order to find QMB ground-states. As a benchmark, we employ this approach against well-known 1D and 2D spin models. Finally, we simulate the $J_1 - J_2$ model, providing the ability to inspect some challenging highly non-trivial models. \\ 

\begin{figure}
\centering
\includegraphics[width=0.4\textwidth]{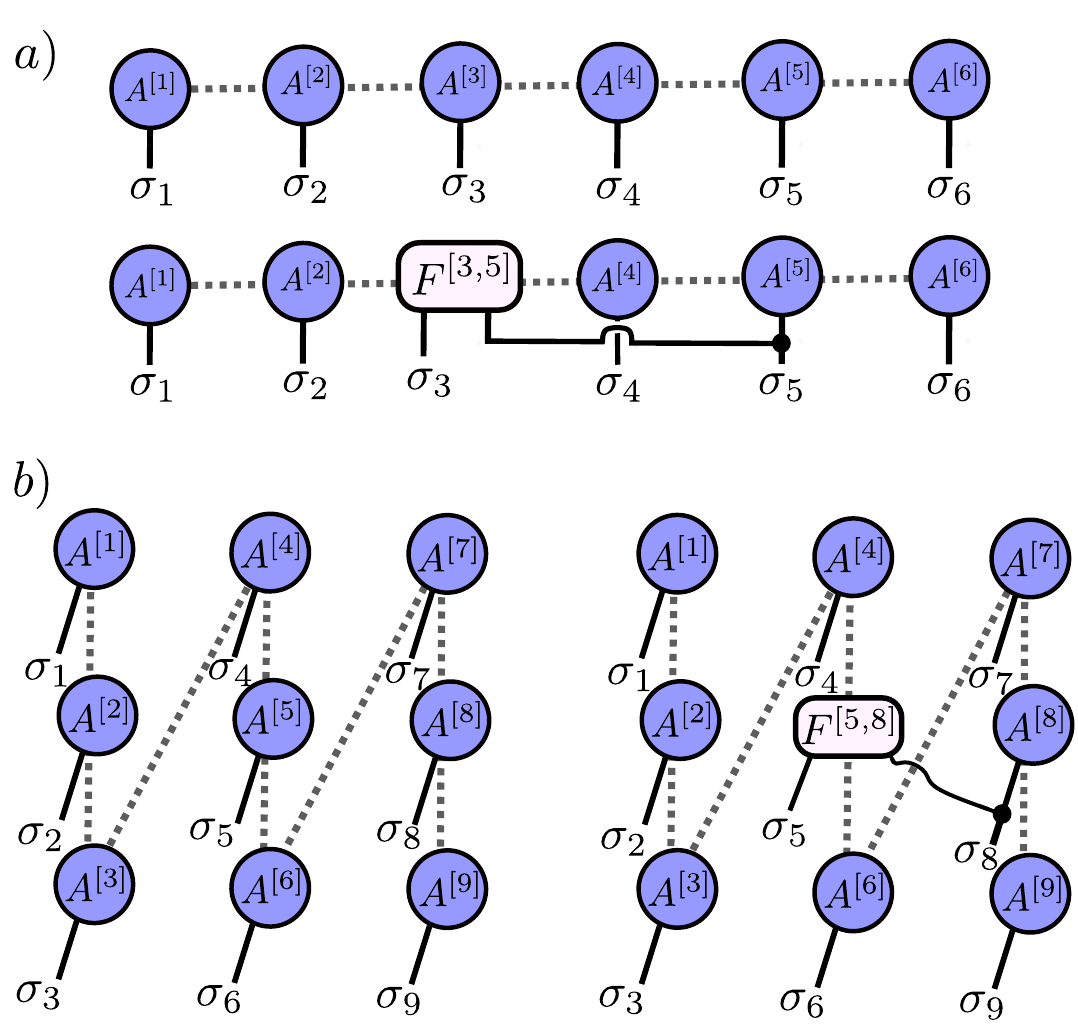}
\caption{Graphical representations of MPS and MPBS applied to 1D ($a$) and 2D ($b$) QMB systems. The $F$ tensors encode correlations between different lattice sites. For illustrative purposes, the pictures represent only one of the terms in which $F$ is involved (see Supplemental Materials). \label{fig:example}}
\end{figure}

\paragraph{Matrix Product Backflow States. ---} 
A state $\ket{\psi}$ of a QMB system consisting of $N$ spin-1/2 variables is fully specified by the complex-valued function $\psi(\pmb{\sigma})= \braket{\pmb{\sigma}|\psi}$, $\pmb{\sigma} \in \{ \pm 1 \}^N$ being the spin projections along the $z$ direction. MPS  \cite{schollwoeck2011} are defined by the functional form  
\begin{equation}\label{eq:mps}
\psi[A](\pmb{\sigma})= A^{[1]}(\sigma_1) A^{[2]}(\sigma_2) ... \,  A^{[N]} (\sigma_N) \, \, ,
\end{equation}
where local tensors $A^{[i]}(\sigma_i)$ 
have one physical index $\sigma_i$ and two auxiliary indices. 
They can be graphically represented as three-legs shapes connected with lines, i.e.\ contracted along auxiliary indices (see Figure \ref{fig:example}) \cite{Penrose1971, schollwoeck2011}. 

\begin{figure*}[ht]
\includegraphics[width=0.95\textwidth]{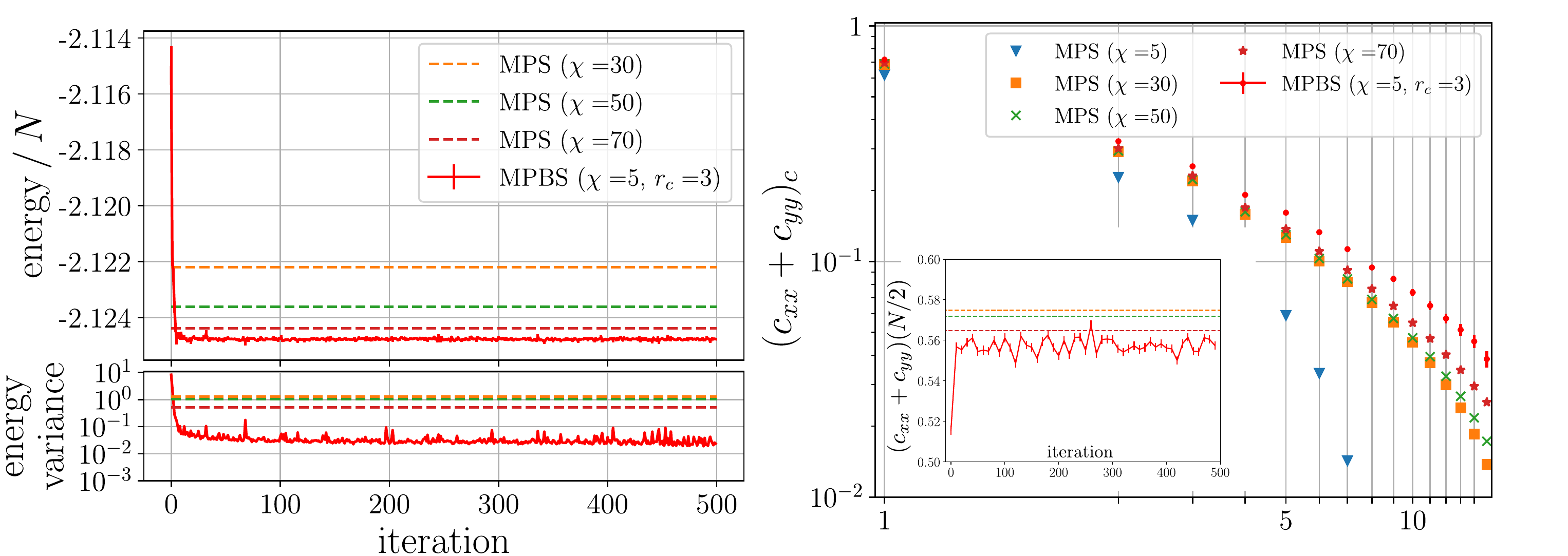}
\caption{MPBS tested of bond dimension $\chi=5$ the modified 1D HS model: energy density convergence (left) and $(c_{xx}+c_{yy})_{\text{c}}$ connected correlator (right). The system size is $N=70$.  
}
\label{fig:HS_1}
\end{figure*}

These indices run from $1$ to a set of integers $\chi_i$, called local bond dimensions, fixing the maximum amount of Entanglement Entropy (EE) which can be encoded by the state \cite{schollwoeck2011}. MPS can provide good approximations of low entangled states, as for instance ground states of local gapped hamiltonians in 1D, for which an area law for EE can be proven \cite{hastings2007}.  On the contrary, MPS cannot efficiently encode a volume law, since this would require an exponentially large value of $\chi$.

In order to overcome these limitations, we introduce a new set of tensors $F^{[i, j]}(\sigma_i, \sigma_j)$ with two physical indices $\sigma_i, \sigma_j$ and two auxiliary indices. These will encode correlations between different lattice sites $i,j$. We propose a new class of wave functions $\psi[A,F](\pmb{\sigma})$ obtained by replacing the MPS \textit{local} tensors $A^{[l]} (\sigma_l)$ as follows:
\begin{equation}\label{eq:newansatzz}
A^{[l]} (\sigma_l) \, \rightarrow \, A^{[l]} (\sigma_l) + \sum_{i_l \neq l} F^{[l,i_l]}(\sigma_l, \sigma_{i_l}) \, \, ,
\end{equation}
which is explicitly depending on the \textit{global} set of quantum numbers $\pmb{\sigma}$. The new wave function can be considered conceptually similar to the well-known backflow wave function in electronic structure theory, which is commonly used to introduce correlations in the mean-field theory by taking the single-particle orbitals act on a configuration-dependent quasi-particle positions \cite{PhysRev.102.1189, schmidt_structure_1981, PhysRevB.78.041101, luo2019}. In our case, the starting point is not a mean-field wave function, but rather an MPS, which can be seen as a systematic and general improvement of the mean-field approximation. 
We thus name this class of variational states Matrix Product Backflow States (MPBS). It is worth mentioning that MPBS wave functions admit a series expansion in increasing powers of $F$, where each term can be formally recast as an MPS with locally larger bond dimension (up to $2^n$ times the original bond dimension, at the order $n$). Examples of first-order ($n\!=\!1$) terms are depicted in Figure \ref{fig:example} (details in Supplemental Material). MPBS with $F$ connected as in Figure \ref{fig:example} (b) will be used in the next sections to simulate 2D systems. They satisfy an area law for EE, since \textit{any} possible grid bipartion cuts a number of auxiliary bonds and/or $F$ tensors that grows linearly with the length of perimeter of the subsystem. Remarkably, it can be easily proven that MPBS ability to encode entanglement can be greater, since with a particular choice of the parameters one can encode a volume law for the EE (see Supplemental Materials). Thus, MPBS can in principle provide good approximations not only of ground-states in 2D, but also of highly entangled QMB states, as for instance time-evolved states after quantum quenches \cite{Alba7947}. From an operative perspective, MPBS naturally suggest a two-steps optimization algorithm: first, the local $A$ tensors are optimized by using the usual MPS machineries; second, the non-local $F$ tensors are optimized by means of VMC techniques. This alternated optimization approaches has the advantage that the starting point of VMC stochastic optimization is not a random point in the parameters space, but rather an already acceptably good approximation of the QMB wave function. Moreover, VMC optimization can further optimize the $A$ tensors as well, thus providing an unrestricted variational search for our ansatz in the last optimization stage. Finally, the MPBS network can be exactly contracted during the Monte-Carlo steps (in contrast to other similar approaches where approximated contraction schemes are employed \cite{LiuPhysRevB.95.195154}), leading to a purely variational scheme.  In the following sections, we will focus on the ground-state search problem, benchmarking the MPBS ansatz on both 1D and 2D models. The numerical results are obtained by means of the two steps optimization algorithm just outlined. The implementation was done in Python by means of \verb|NetKet| \cite{carleo2019,netket,vicentini2021netket}, a package providing machine learning and automatic differentiation methods for QMB systems (see Supplemental Materials for technical details). 

\begin{figure*}[!ht]
\includegraphics[width=0.95\textwidth]{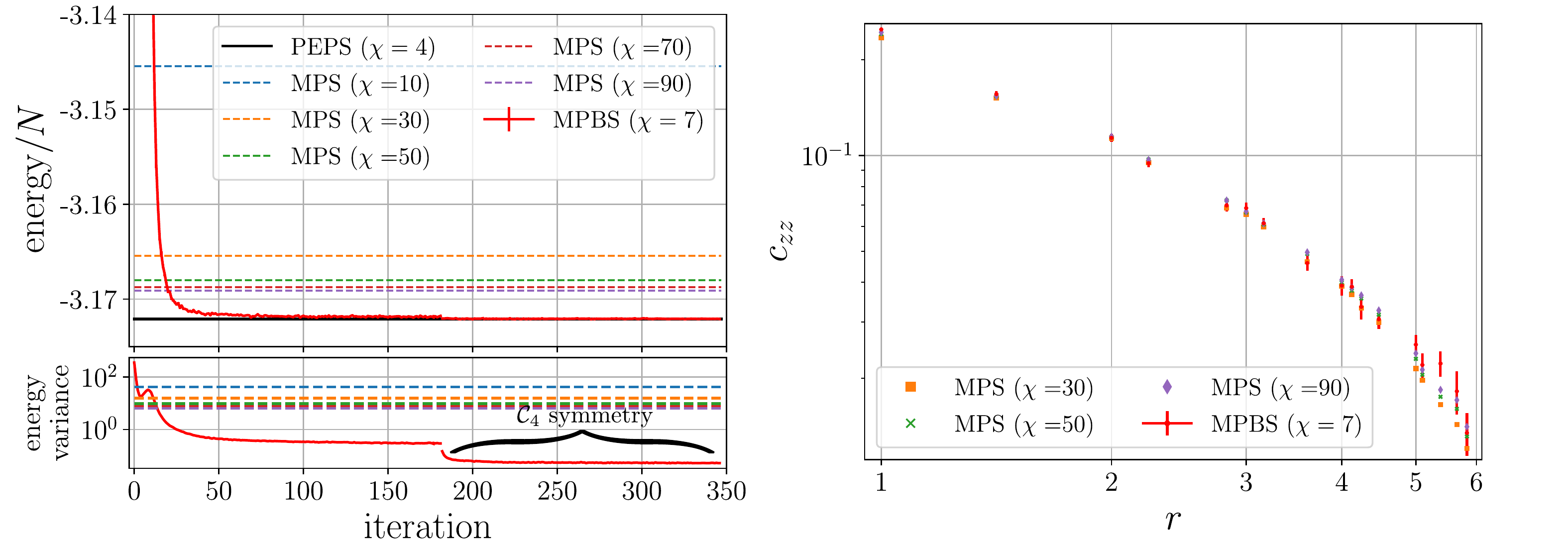}
\caption{MPBS of bond dimension $\chi=7$ tested with the 2D Ising model on a square lattice ($N_x=11$, $N_y=11$): energy density convergence (left) and $c_zz$ correlator (right).PEPS result is taken from
\cite{Lubasch_2014}.}
\label{fig:im_2}
\end{figure*}

\begin{figure*}[!ht]
\includegraphics[width=0.95\textwidth]{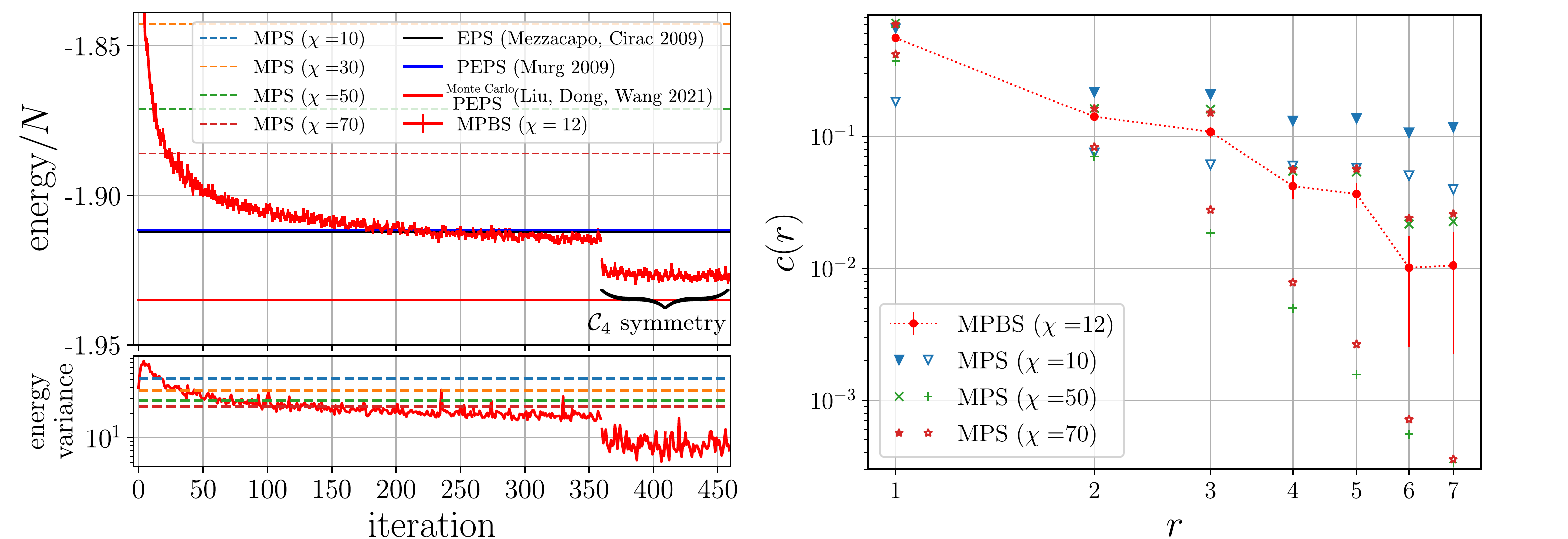}
\caption{MPBS of bond dimension $\chi=12$ tested with the $J_1-J_2$ model on a square lattice ($N_x=8$, $N_y=8$). PEPS and EPS results are taken from Ref.\cite{Mezzacapo_2009}. In the right plot, filled (empty) markers represent DMRG results for $c_{ver}(r)$ ($c_{hor}(r)$), whereas red points are MPBS results.}
\label{fig:im_3}
\end{figure*}

\paragraph{Modified Haldane-Shastry model. ---} 
First, we apply MPBS to a 1D quantum spin chain with Periodic Boundary Conditions (PBC). In particular, we consider the following modified Haldane-Shastry (HS) model $\mathcal{H}_{HS} = \sum_{j < i} \big(1 / \widetilde{d}_{ij} \big)^2 \big( - \sigma^x_i \sigma^x_j - \sigma^y_i \sigma^y_j + \sigma^z_i \sigma^z_j \big)$, 
where $\widetilde{d}_{ij} = N/\pi \cdot \sin (\pi/N |i\!-\!j|)$.
This model is known to be particularly challenging for standard DMRG, as it shows power-law scaling in the ground-state EE~\cite{deng2017}. To use our optimization scheme, we adapt the MPBS ansatz in order to explicitly realize translational invariance. This is achieved by adding an extra auxiliary index, connecting the first and the last site, as well as by taking the $A$ tensors independent from the site $i$. Also, we set $F^{[i,j]}$ to be dependent only on the distance $d_{ij}=\min(|i-j|,  N \! -\!  |i \! - \! j| )$ between the two connected sites. We also introduced a cut-off $r_c$ setting the maximum distance between sites for which the $F$ tensors are non-zero (i.e.\ $F^{[d_{ij}]}(\sigma_i, \! \sigma_j) = 0$ if $ d_{ij} \! > \! r_c$). Due to translational invariance and the imposed cutoff, the number of variational parameters of the ansatz is independent of the system size $N$, resulting in a reduced computational cost for the Monte-Carlo simulation. In the first optimization stage, we write the HS hamiltonian as a Matrix Product Operator (MPO) and use standard two-sites DMRG~\cite{schollwoeck2011} to get the optimized $A$ tensors. In the second stage, a VMC optimization of the $F$ tensors is realized, adopting the Stochastic Reconfiguration \cite{Sorella_1998} \emph{natural} gradient descent approach. 
Since $\mathcal{H}_{HS}$ commute with the total $z-$magnetization $\Sigma^z = \sum_{i=1}^N \sigma_i^z$ and the parity operator $P=\sigma_1^x \sigma_2^x ... \, \sigma_N^x$, we restrict the ground-state search to the $\Sigma^z =0$ sector of the Hilbert space. 
In Figure \ref{fig:HS_1}, we show some selected results, obtained with a relatively small value of the MPBS bond dimension ($\chi=5$) and $r_c=3$. First subplot shows the expectation value and the variance of the energy, tracked during the VMC optimization (red lines). Dotted lines represent DMRG energies/variances for increasing values of the bond dimension. After less than $10^2$ VMC optimization steps the MPBS energy reach energy values smaller than the DMRG energy obtained with the larger value of $\chi$ ($\chi=70$). Let us remark that number of parameters this MPS is much larger than the number of them parameters of our ansatz, meaning that MPBS provide good approximations of the true QMB ground-state. Moreover, we also got substantially better results in terms of energy variance. 
Second subplot in Figure \ref{fig:HS_1} shows the two-points connected correlator $\big(c_{xx}+c_{yy}\big)_{\text{c}}$, computed by taking average of $\braket{\sigma_i^x \sigma_{i+r}^x} + \braket{\sigma_i^y \sigma_{i+r}^y}$ over $i$ and then subtracting the square of the average $x$ and $y$ magnetizations. Red points represent estimations obtained at the end of the VMC optimization, whereas other points are DMRG results. These seem to converge to VMC values, when increasing the bond dimension $\chi$. In the inset it is shown the correlator $c_{xx}(N/2)+c_{yy}(N/2)$ as estimated during the Monte-Carlo iterations. The convergence appears to be fast. Other applications of MPBS ansatz to 1D systems are reported in Supplemental Materials.

\paragraph{Two dimensional Ising model. ---} 
To corroborate the flexibility of MPBS in describing higher dimensional systems, we now start analyzing 2D QMB models living on a square lattice of size $N_x \times N_y$ with Open Boundary Conditions (OBC). A simple way to adapt MPS to the description of such a system is to order the sites of the grid following a one-dimensional ``snaking" path connecting all the sites (see Figure \ref{fig:example} $b)$) \cite{cirac2018}. Others 2D to 1D mappings have been also studied \cite{cataldi2021}, leading to increased numerical precision but not to a significant improvement in the codification of entanglement in 2D systems. The main issue is that, since area law in 2D implies that EE grows linearly with the length of the subsystem perimeter, any MPS cannot describe efficiently typical ground-states of 2D hamiltonians. As a possible improvement, we propose to arrange the MPBS ansatz in order to codify correlations between sites which are adjacent in the 2D geometry but which are placed at distance $N_y$ along the 1D snaking path. This can be done by setting the $F^{[i]}(\sigma_i, \sigma_{j})$ matrices different from zero in the cases in which $j \! = \! i\pm N_y$, where we label the lattice sites with a single integer $i \! = \! 1,2,...\, N$. As already mentioned, an MPBS of this kind can encode the area law for the EE and, at least for a particular choice of the parameters, the volume law  (see Supplemental Material). To benchmark the efficacy of MPBS in simulating 2D systems, we consider the following Ising hamiltonian \ $\mathcal{H} =  - \sum_{\braket{\pmb{i},\pmb{j}} } \sigma_{\pmb{i}}^z  \sigma_{\pmb{j}}^z + h \sum_{\pmb{i}} \sigma_{\pmb{i}}^x$ on a lattice of dimension $N_x \! = \!N_y \!= \!11$. In Figure \ref{fig:im_2}, we show the results of an MPBS optimization, runned with bond dimension $\chi=5$ and transverse field $h=3.0$, close to the quantum critical point of the system $h_c \simeq 3.044$ \cite{PhysRevE.66.066110}.
These results are compared with DMRG findings at different bond dimensions and with the energy value obtained by Lubasch and others by means of PEPS \cite{Lubasch_2014}.
As in the previous case, MPBS with extremely small bond dimension leads, after $\approx 100$ VMC optimization iterations, to results significantly better than DMRG, both in terms of energy density and energy variance. Since the system has rotational symmetry, during the last $\simeq 150$ Monte-Carlo iterations we explicitly symmetrize the MPBS with respect to the $\mathcal{C}_4$ group of fourfold rotations. To do this, we consider the following modified wave function $\psi'[A,F](\pmb{\sigma}) = \sum_{k=0}^3 \psi[A,F](R^k\pmb{\sigma})$, where $R$ is a rotation of $\pi/2$ of the spin configuration. This results in a further improvement of the energy and energy variance. The value of energy density we find at the end of the optimization is $\braket{\mathcal{H}}/N = -3.17208(1)$. In the second subplot, we show the correlator $c_{zz}(r)= 1/N_r \cdot \sum_{\pmb{r}, |\pmb{r}|=r } \braket{\sigma^z_{\pmb{i}_c} \sigma^z_{\pmb{i}_c + \pmb{r}}}$, where $\pmb{i}_c$ indicates the central site of the grid and $N_r$ is the number of sites placed at distance $r$ from this. MPBS points seem to be in good agreement with the trend of DMRG results for increasing bond dimension.

\paragraph{Two dimensional $J_1-J_2$ model. ---} Finally we consider the anti-ferromagnetic $J_1-J_2$ model, with hamiltonian $\mathcal{H} =  J_1 \sum_{\braket{\pmb{i},\pmb{j}} } \sigma_{\pmb{i}} \cdot  \sigma_{\pmb{j}} + J_2 \sum_{ \braket{\braket{\pmb{i},\pmb{j}}} } \sigma_{\pmb{i}} \cdot  \sigma_{\pmb{j}}$, 
where the first (second) sum is on first (second) nearest neighbors couples of sites. This is a prototypical frustrated magnetic system. Despite active research in the past decades  \cite{PhysRevB.88.060402, PhysRevLett.84.3173, PhysRevB.100.125124, PhysRevB.98.241109}, the nature of the ground-state around the point of maximum frustration $J_2/J_1 \! = \! 0.5$ remains unclear. We address the problem Hamiltonian by means of MPBS arranged as in the previous paragraph and also adding $F$ tensors connecting second nearest neighbors sites (see Supplemental Materials for details). As in the HS model, we reduce the simulation to the zero magnetization sector. In Figure \ref{fig:im_3} we show some selected results obtained with a system of size $N_x \! = \!N_y \! = \! 8$, OBC and $J_1=1, J_2=0.5$. 
After $\simeq 350$ VMC optimization iterations, we apply $\mathcal{C}_4$ wave function symmetrization. We compare our results with the EPS and PEPS results reported in \cite{Mezzacapo_2009} and with Monte-Carlo optimized PEPS results reported in \cite{PhysRevB.98.241109}. The final energy density of our simulation is $\braket{\mathcal{H}}/N=-1.9273(9)$ and is lower than both values reported in \cite{Mezzacapo_2009}, whereas it is about $\approx 7 \! \cdot \! 10^{-3}$ greater than the value reported in \cite{PhysRevB.98.241109}. It should be however remarked that the value in \cite{PhysRevB.98.241109} is not strictly variational, because of the approximate contraction scheme adopted for PEPS. 
Finally, we measure some relevant observables as the correlators $c_{\text{ver}}(r) = 1/N_x \cdot \sum_{j} \braket{\pmb{\sigma}_{1,j} \cdot \pmb{\sigma}_{1+r, j}}$ and $c_{\text{hor}}(r) = 1/N_y \cdot \sum_{i} \braket{\pmb{\sigma}_{i, 1} \cdot \pmb{\sigma}_{i, 1+r}}$, which are shown in the second half of Figure \ref{fig:im_3}. These are respectively the average spin-spin correlators along the columns and the rows of the grid. Since the wave function $\psi'[A,F](\pmb{\sigma})$ is symmetric under rotation of $\pi/2$, we always find values for these correlators 
compatible within the uncertainty bars. On the contrary, DMRG results 
show that MPS are unable to encode power-law decaying correlations along the horizontal direction. We also measure the structure factor $S^2(\pmb{q})=1/(N(N+2)) \cdot \sum_{\pmb{i}, \pmb{j}} \braket{\pmb{\sigma}_{\pmb{i}} \cdot \pmb{\sigma}_{\pmb{j}}}e^{-i \pmb{q} \cdot (\pmb{i} - \pmb{j})}$ for different pitch vectors $\pmb{q}$. We find 
$S^2(0,\pi) \simeq 3.19(5) \cdot 10^{-2}$
and $S^2(\pi,\pi) \simeq 0.241(3)$. The latter corresponds to the Néel order
parameter. Both values are compatible with similar findings in \cite{PhysRevB.100.125124}. We also obtain 
$S^2(0,0)=1.3(2) \cdot 10^{-4}$, which is consistent with the expectation that the 
$J_1-J_2$ ground state is in a singlet under SU(2) global symmetry.

\paragraph{Conclusions and outlook. ---}
We have introduced a novel variational {\it ansatz}
which exploit state-of-the-art numerical techniques based on Tensor Networks and automatically differentiable Variational Monte Carlo. The new many-body wave function encodes area law entanglement for high dimensional QMB systems. The efficiency of MPBS allows to study challenging 2D models, encoding accurate long-range correlations and going beyond the {\it standard} PEPS and DMRG ansatze. The MPBS structure takes its root from the usual MPS, whose descriptive power is augmented by introducing a new class of long-ranged tensors. The new arranged network is well suited for a two-step optimization scheme, i.e.\ DMRG followed by VMC. The second step takes tremendous advantages from using an already reasonably good initialization of the QMB wave function and also to the use of an exact contraction scheme. The results presented in this work regarding some well-known models, as 1D and 2D Ising model and 1D Haldane-Shastry model, provide evidences that MPBS constitute a good ansatz to approximate ground-states of QMB systems both in 1D and 2D, via a purely variational approach. It can also be employed to study highly non-trivial systems, as the 2D $J_1 - J_2$ frustrated model. 
The final optimized wave function can be easily used to compute useful observables, as for instance spin-spin correlators. 
Significantly, the outlined two-steps method can be in principle applied also to the real-time dynamics problems. In this case, Time-Depedent Variational Principle (TDVP) \cite{Vanderstraeten2019} followed by time-dependent VMC can be used.

\bibliography{bib}

\newpage
\onecolumngrid
\appendix
\appendixtitleon
\appendixtitletocon

\begin{appendices}
\section{SUPPLEMENTARY MATERIALS}

\section{Implementation}
All the numerical simulations were performed on SISSA's Ulysses cluster. The code implementing the MPBS ansatz and its optimization was fully written in Python. The first step of the alghoritm, i.e.\ DMRG, has been written by means of the Python package \verb|mpnum|  \cite{Suess2017}. In all the shown simulations, a two-site DMRG approach with initial random MPS was used. The second step, i.e.\ the VMC code, has been written with  version 3 of \verb|NetKet| \cite{carleo2019, vicentini2021netket, netket}. \verb|NetKet| is an open-source project delivering cutting-edge methods for the study of many-body quantum systems with artificial neural networks and machine learning techniques. It is a Python library built on JAX \cite{jax2018}, a package providing Automatic Differentiation routines. \verb|NetKet| offers several methods to define custom models. The MPBS model was implemented by using the Flax Linen API framework \cite{flax2020github}. The VMC optimization was written in NetKet, by using the Stochasitc-Reconfiguration (SR) preconditioner (also known as Natural Gradient Descent, in the machine learning literature) \cite{Sorella_1998}. As sampler routines, we used or the standard local Metropolis algorithm or, in the cases in which the total magnetization was keep fixed, a variant in which new system configurations are generated by flipping couples of spins, thus preserving the total $z-$magnetization (see \cite{netket} for details). In all the shown simulations, $4$ Markov chains runned in parallel on a single process, while between $2$ and $4$ cluster nodes were used. The number of samples per Monte-Carlo step was chosen between $10^3$ to $10^4$, depending on the simulation. At the beginning of the VMC optimization, $F$ tensors were initialized as follows 
\begin{equation}
  F^{[i,j]}(\sigma_i, \! \sigma_j) = \epsilon ( \mathbb{1}_{\chi \times \chi} + \sigma \zeta_{\chi \times \chi} )  \, \, , 
\end{equation}
where $\mathbb{1}_{\chi \times \chi}$ is the identity matrix of dimension $\chi$,  $\zeta_{\chi \times \chi}$ is a random matrix with normally distributed entries and $0 < \sigma, \epsilon \ll 1$ are (small) real numbers (usually we set to $\epsilon = 0.01$ or $0.005$ and $\sigma=0.1$).

\subsection{Other numerical results}
In this section we provide some additional numerical results, obtained with the MPBS ground-state search scheme. 

\subsubsection{Modified Haldane-Shastry model}



\begin{figure}[ht!]
\centering
\includegraphics[width=0.5\textwidth]{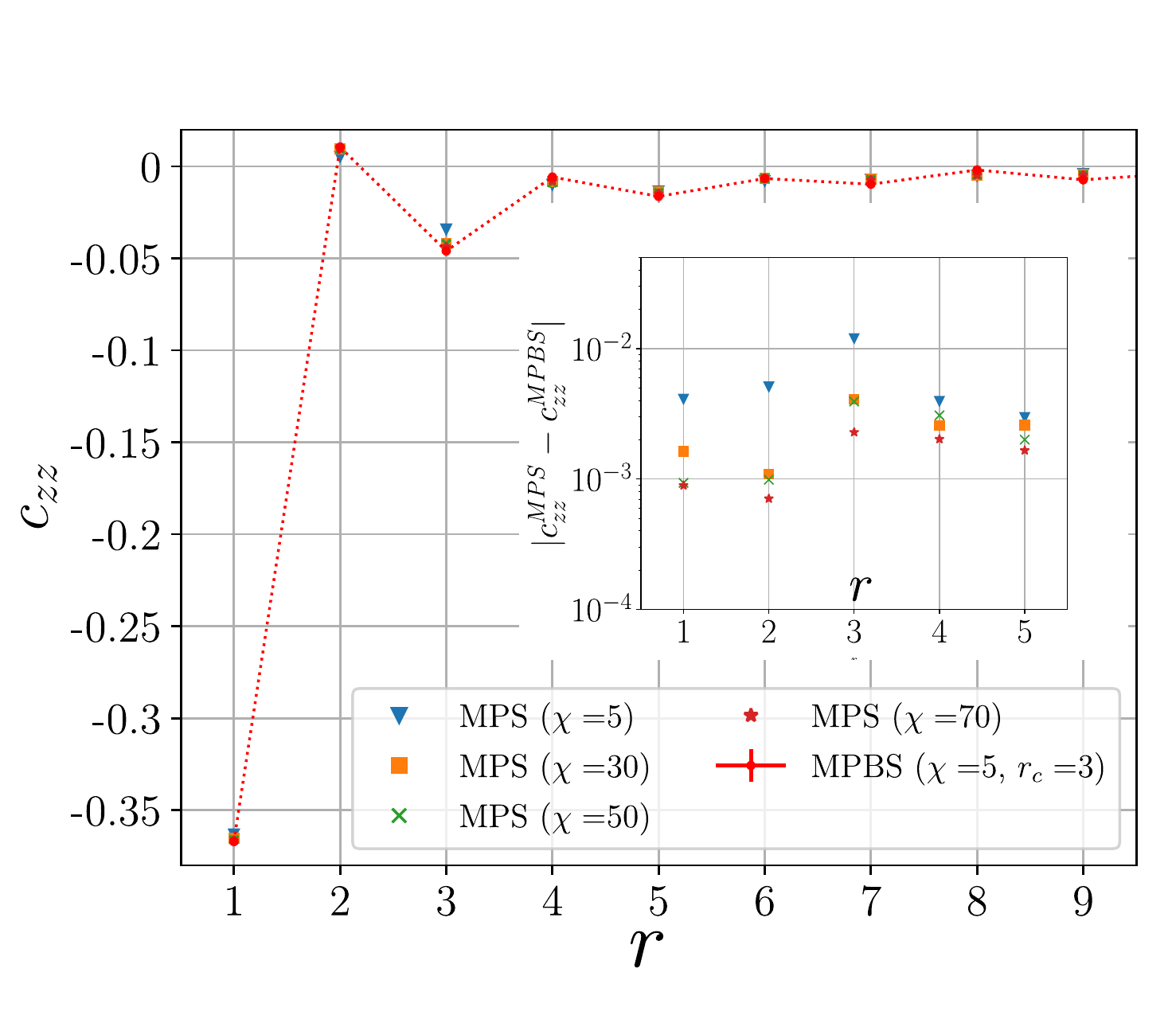}
\caption{The spin-spin correlator $c_{zz}(r)$ of the 1D HS Model ($N=70$). The inset represent $|c_{zz}^{MPS}(r) - c_{zz}^{MBPS}(r)|$. 
\label{fig:HS_2}}
\end{figure}

Figure \ref{fig:HS_2} shows the two-points correlators $c_{zz}(r)$ as estimated at the end of the VMC optimization (red points), for the same simulation shown in the main text. Since translational invariance was explicitly realized by our ansatz, a spatial average over was taken, i.e. we considered $c_{zz}(r) = 1/N \sum_i \braket{ \sigma^z_i \sigma^z_{i+r} }$.  Results are substantially in agreement with the ones displayed in \cite{deng2017}. The other points represent DMRG results for increasing values of the bond dimension.
In the inset, we plotted the absolute value of the difference between the DMRG correlators and the estimated VMC correlators. By increasing $\chi$, the DMRG correlators goes monotonically to the VMC values.

\subsubsection{1D Ising model}

\begin{figure}[ht!]
\centering
\includegraphics[width=0.55\textwidth]{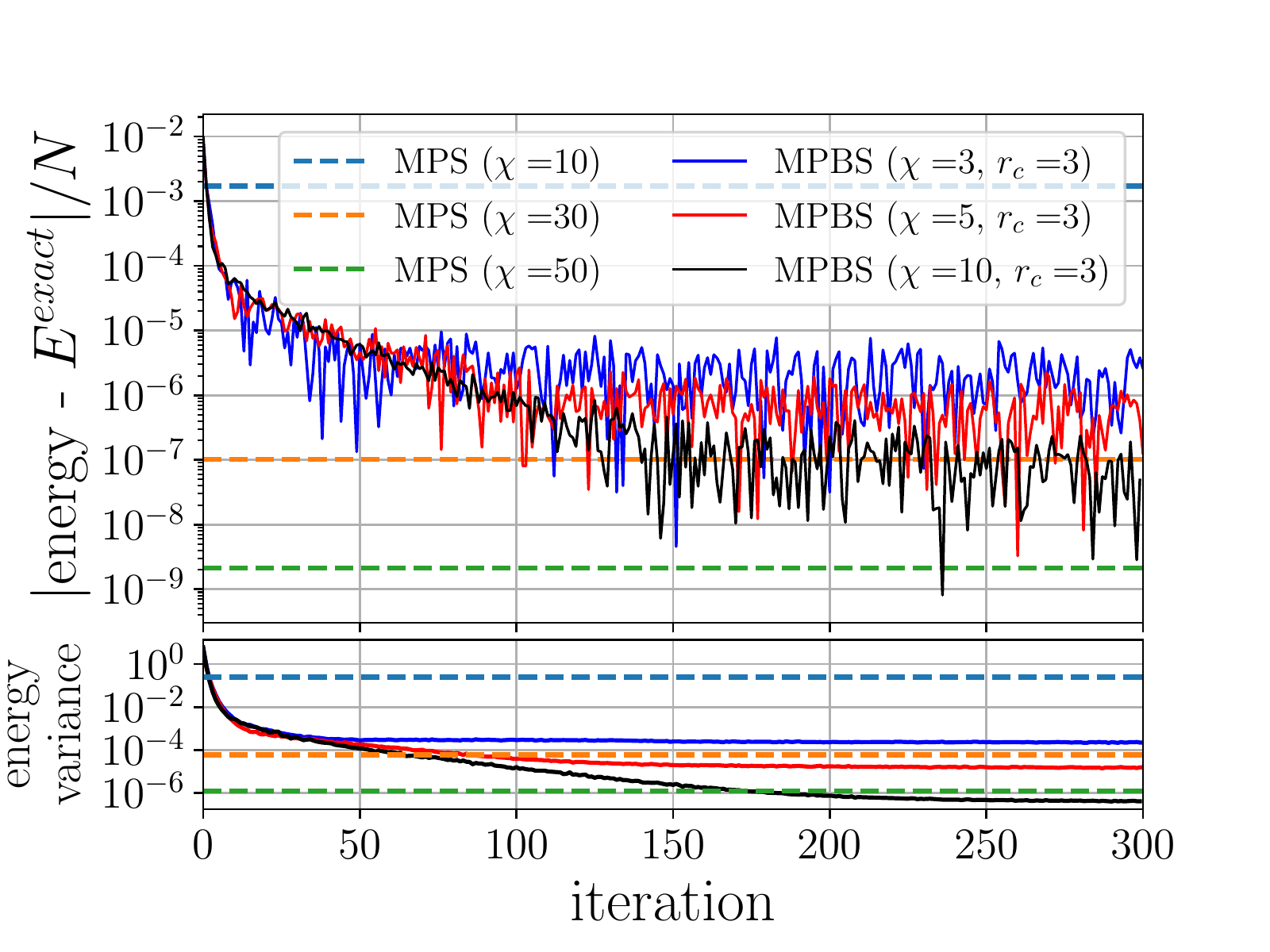}
\caption{MPBS tested on the TFPIC at $h=1.1$. The system size used was $N=100$. 
\label{fig:is_1}}
\end{figure}

We also performed simulations with the standard Transverse-Field Periodic Ising Chain (TFPIC), with nearest-neighbours interactions and PBC. Our hamiltonian was $\mathcal{H} = - \sum_{i=1}^{N} \sigma^z_{i}  \sigma^z_{i+1}  - h \sum_{i=1}^N \sigma^x_i$, 
with $\sigma^z_{N+1}=\sigma^z_{1}$. To use DMRG for this hamiltonian, we developed an appropriate trick to implement PBCs (see last section of Supplemental Materials). As far as we know, this trick is not explicitly present in literature. In Figure \ref{fig:is_1}, we show some results of simulations obtained with a system of size $N=100$ and $h=1.1$ (i.e.\ close to the quantum critical point at $h=1.0$). The MPBS cut-off was $r_c=3$ and we considered different bond dimensions $\chi=3,5,10$ Since the TFPIC can be solved exactly by means of the Jordan-Wigner transformation \cite{mbeng2020quantum}, we computed  deviations of the variational energies per site from exact energy per site. It is known that this model does not present any particular difficulty for standard two-sites DMRG aloghoritm and indeed with a bond dimension $\chi=30$ DMRG reaches a deviation of order $10^{-7}$ (see the orange line in Figure \ref{fig:is_1}). We reached deviations of the same order of magnitude by means of 
MPBS with $\chi=10$, optimized with DMRG followed by $\simeq 300$ Monte-Carlo iterations. Remarkably, MPBS were able to get values of the energy variance even smaller of the ones obtained by DMRG with MPS and $\chi=10,30,50$.

\subsubsection{2D Ising model}
\begin{center}
\begin{figure}[ht!]
\centering
\includegraphics[width=0.5\textwidth]{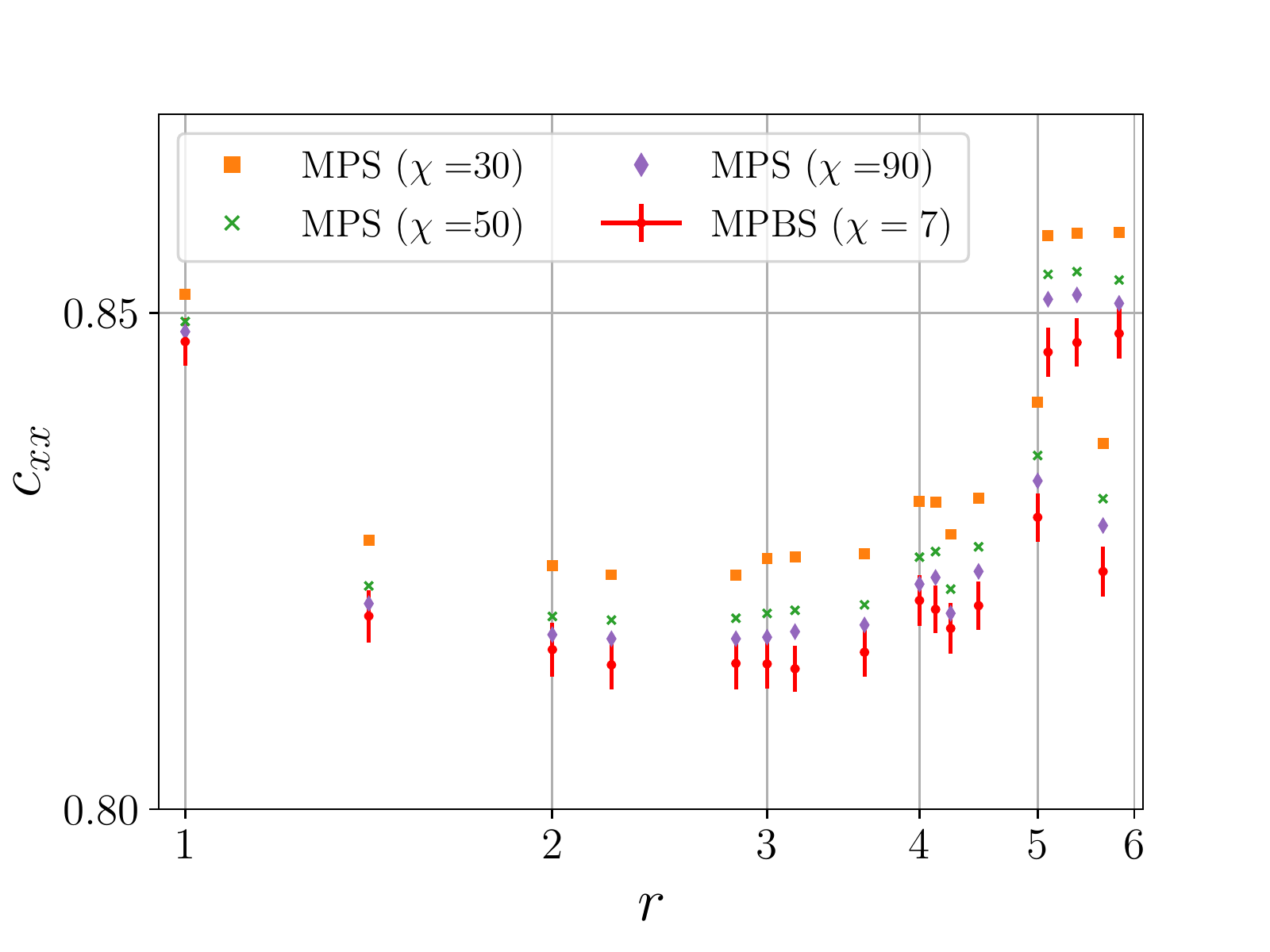}
\caption{The spin-spin correlator $c_{xx}(r)$ of the 2D Ising model.
\label{fig:cxx_ising_2d}}
\end{figure}
\end{center}

Figure \ref{fig:cxx_ising_2d} shows the correlator $c_{xx}(r)= 1/N_r \cdot \sum_{\pmb{r}, |\pmb{r}|=r } \braket{\sigma^z_{\pmb{i}_c} \sigma^z_{\pmb{i}_c + \pmb{r}}}$, where $\pmb{i}_c$ indicates the central site of the $11 \times 11$ grid and $N_r$ is the number of sites placed at distance $r$ from this. As usual, this observable were measured at the end of the two steps optimization. The simulation is the same shown in the main text. DMRG points seems to flow into the VMC results by increasing the bond dimension $\chi$ of the MPS. \\

\begin{center}
\begin{figure}[ht!]
\centering
\includegraphics[width=0.45\textwidth]{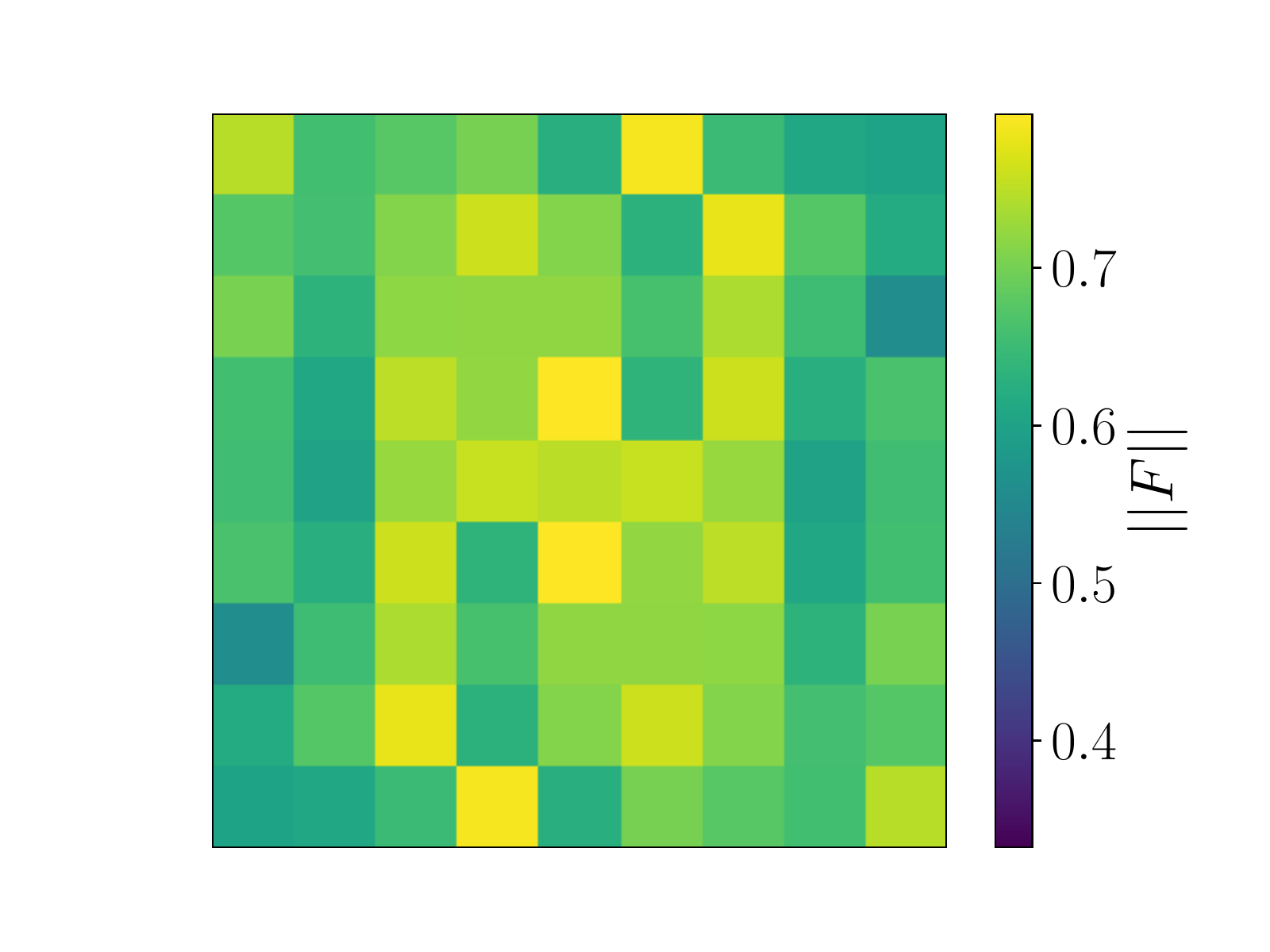}
\caption{Site by site norm of the $F$ tensors, as computed at the end of the optimization. Only the $9 \times 9$ inner part of the lattice in considered.
\label{fig:F_norms}}
\end{figure}
\end{center}

In the colour plot in Figure \ref{fig:F_norms}, instead we show the 
norms of the $F$ tensors obtained at the final step of the VMC optimization. Norms are defined as $||F||=\text{tr}(F^{\dag} F)$, where $F$ and $F^{\dag}$ are contracted on all their indices (i.e.\ two physical indices and two auxiliary indices). Site by site, we also performed an average between the norms of the two tensors $F(i, i\pm N_y)$. \\ 

\vspace{-12 mm}

\vspace{-14 mm}

\begin{figure}[ht!]
\centering
\includegraphics[width=0.5\textwidth]{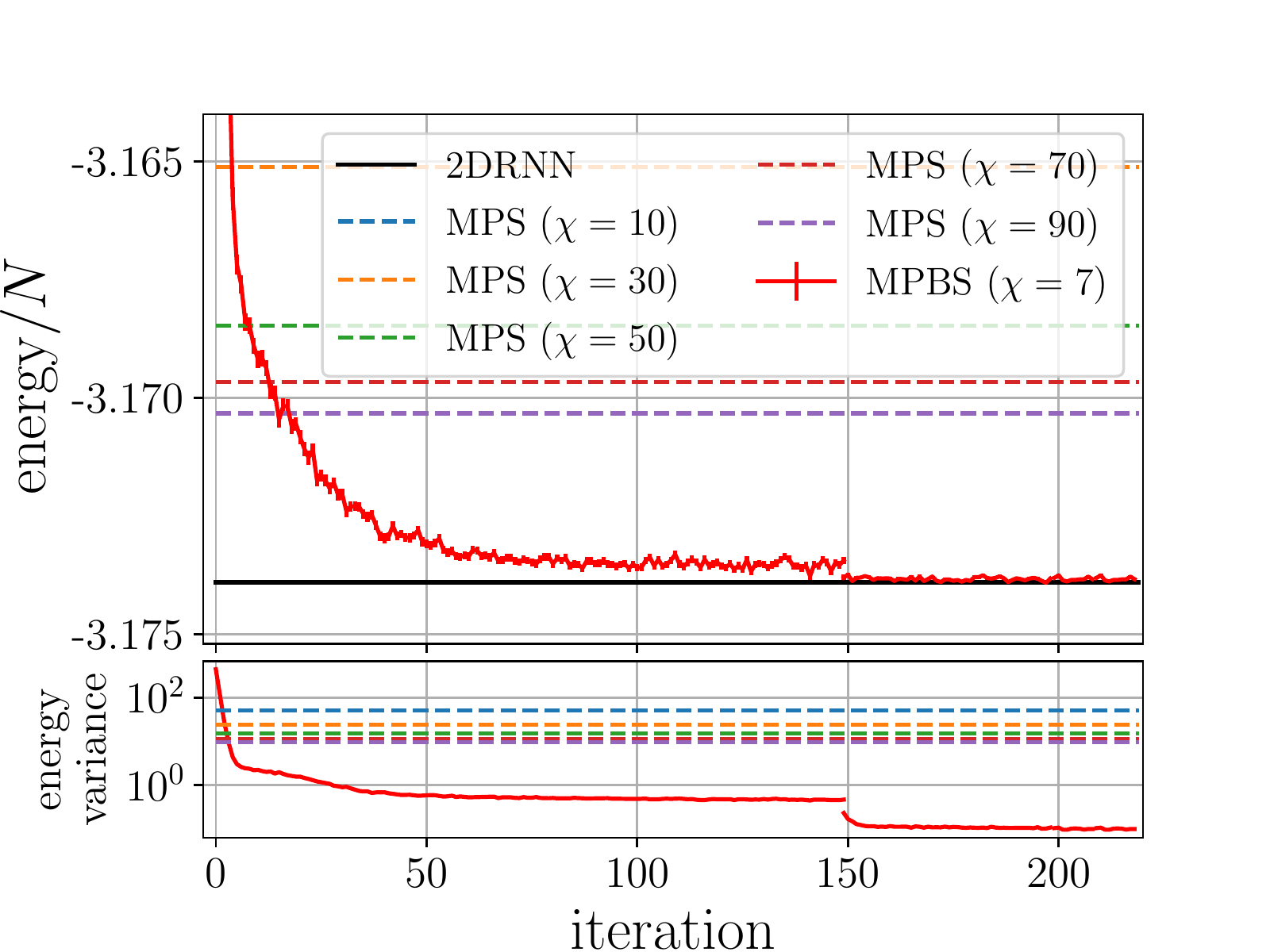}
\caption{MPBS of bond dimension $\chi=7$ tested with the 2D Ising model on a square lattice ($N_x = 12, N_y = 12$)
\label{fig:is_12}}
\end{figure}

In order to compare MPBS performance with the extremely accurate results obtained in Ref. \cite{PhysRevResearch.2.023358} by means of Recurrent Neural Network (RNN) wave functions, we also performed a simulation of an Ising model of size $N_x = N_y =12$. We set again $h=3.0$ and we used $\mathcal{C}_4$ symmetryzation trick in the last $\approx 70$ Monte-Carlo iterations. Results are shown in Figure \ref{fig:is_12}. The energy density we found at the end of the optimization was $\braket{\mathcal{H}}/N = -3.17385(1)$, whereas in Ref. \cite{PhysRevResearch.2.023358} the authors obtain $\braket{\mathcal{H}}/N =-3.1739018(2)$ with a single layer of 2D RNN with 100 memory units.

\subsubsection{$J_1 - J_2$ model}
\begin{figure}[ht!]
\centering
\includegraphics[width=0.2\textwidth]{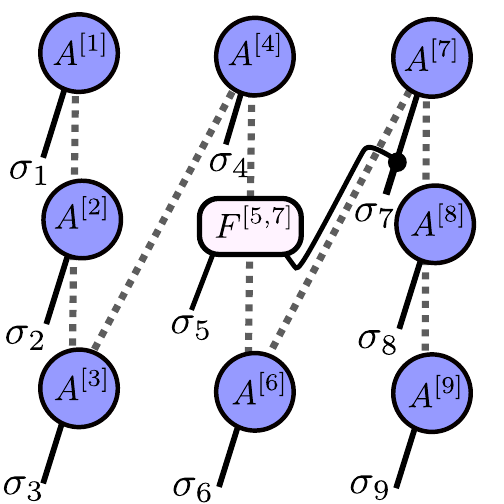}
\caption{A generic second nearest neighbors connections in the two dimensional MPBS. \label{fig:secondivicini}}
\end{figure}

In order to simulate the highly non-trivial ground state of the 2D $J_1-J_2$ model, in addition to the $F$ tensors connecting nearest neighbors sites, we also considered new tensors connecting second nearest neighbors sites. These are graphically represented in Figure \ref{fig:secondivicini} and correspond to term as $F^{[i]}(\sigma_i, \sigma_{i \pm N_y \pm 1})$, if we label the sites with a single integer running along the snaking path.

\subsection{Decomposing the MPBS as a sum of MPS}

As mentioned, the MPBS wavefunction can be expanded in a series of terms at different orders in powers of $F$: $\psi[A, F](\pmb{\sigma}) = \sum_{n=0}^{N} \psi^{(n)}[A, F](\pmb{\sigma})$. The first ones are 

\small
\begin{align}\label{eq:newansatzp}
\begin{split}
    \psi[A, F](\pmb{\sigma}) &= 
\underbrace{A^{[1]}_{\alpha_1}(\sigma_1)  ... \,  A^{[N]}_{\alpha_N}(\sigma_N)}_{\psi^{(0)}[A](\pmb{\sigma})}
+ \\ &+ \underbrace{\sum_{i \neq j} \bigg( A^{[1]}_{\alpha_1}(\sigma_1) ... \, F_{\alpha_{i-1} \alpha_i}^{[i, j]}(\sigma_i, \sigma_j) ... \, A^{[N]}_{\alpha_N}(\sigma_N) \bigg)}_{\psi^{(1)}[A, F](\pmb{\sigma})} \\ &+ \, \, o \big(F^2 \big) \, \, .
\end{split}
\end{align}
\normalsize

\begin{figure}[ht!]
\centering
\includegraphics[width=1.0\textwidth]{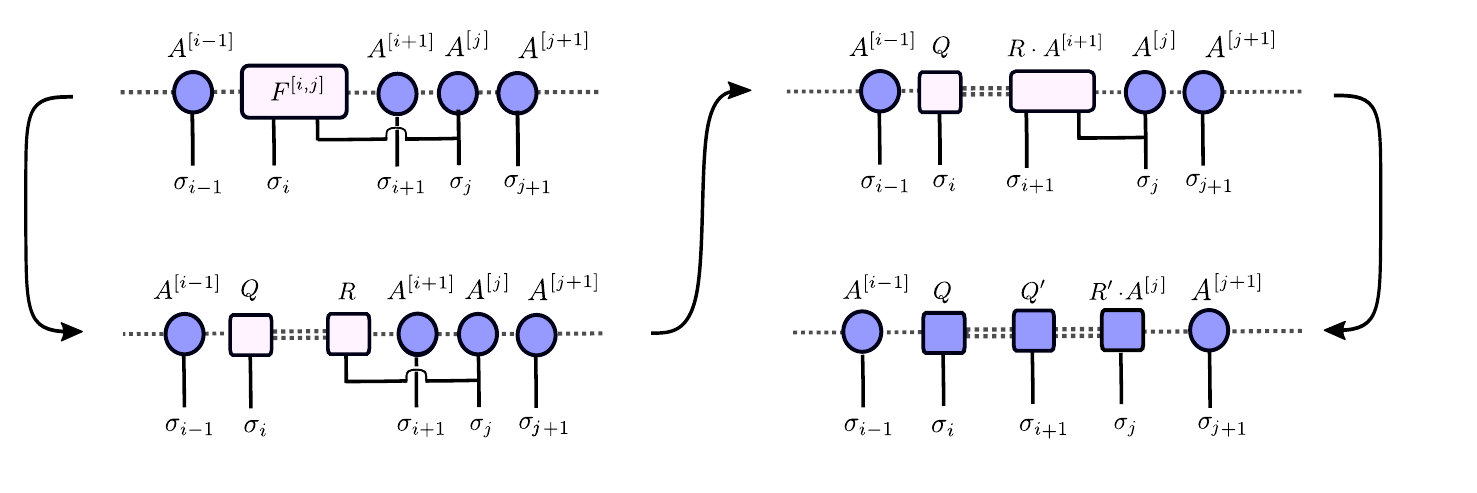}
\caption{decomposition of one term of $\psi^{(1)}[A,F]$ as MPS. We arbitrarily set $j=i+2$. Double dotted lines represent auxiliary bonds/indices with local bond dimension $2 \chi$, whereas dotted lines have bond dimension $\chi$. 
\label{fig:exampledecomposition}}
\end{figure}

\begin{figure}[ht!]
\centering
\includegraphics[width=0.45\textwidth]{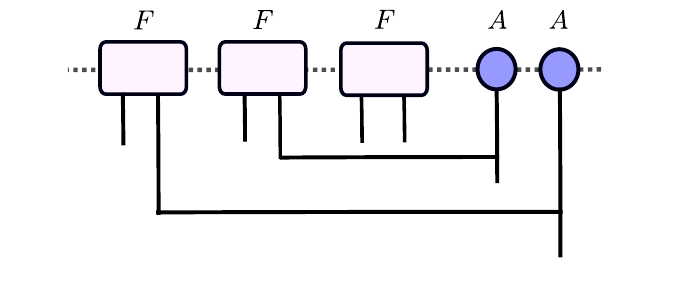}
\caption{A term contained in $\psi^{(n)}$ ($n=3$) with $n$ overlapping long-range connections.  
\label{fig:order_n}}
\end{figure}

The zero order term $\psi^{(0)}$ is just the MPS wave function, whereas the first order term $\psi^{(1)}$ consists of $N$ pieces as the one graphically represented in Figure \ref{fig:example} with the usual TN graphical notation \cite{Penrose1971, schollwoeck2011}. In a generic term $\psi^{(n)}$ of the expansion, we would find $n$ $F$ tensors, each connecting a couple of physical indices. $\psi^{(N)}$ can give rise to the well-known Jastrow wave function as a particular case (when setting $\chi=1$, $A=0$)
\begin{equation}
\psi_{\text{Jastrow}}(\pmb{\sigma})= \prod_{i<j} F^{[i,j]}(\sigma_i, \sigma_j)  \, \, .
\end{equation}
Let us now focus on the linear term. We can reshape the tensor $F$ into a square matrix $\tilde{F}$ of dimension $2 \chi$
\begin{equation*}
F_{\alpha_{i-1} \alpha_i}^{[i, j]}(\sigma_i, \sigma_j) = \tilde{F}_{(\alpha_{i-1} \sigma_i), (\alpha_i \sigma_j)}=\tilde{F}_{aa'} \, \, ,
\end{equation*}
where $\alpha_{i-1}, \alpha_i$ are the two original auxiliary indices. We introduced the indices $a,a'=1,2... \, 2 \chi$. By applying a QR-decomposition, we obtain 
\begin{equation*}
\tilde{F}_{aa'}=Q_{ab} R_{ba'}=Q_{ \alpha_{i-1} b}(\sigma_i) R_{b \alpha_i}(\sigma_j) \, \, .
\end{equation*}
In order to simplify the notation, let us consider the particular case 
$j=i+2$ (the generalization to the general case will be trivial). By applying another QR-decomposition, we get
\begin{align*}
\begin{split}
&... \, Q_{ \alpha_{i-1} b}(\sigma_i) R_{b \alpha_i}(\sigma_{i+2}) A_{\alpha_i \alpha_{i+1}}^{[i+1]}(\sigma_{i+1}) A_{\alpha_{i+1} \alpha_{i+2}}^{[i+2]}(\sigma_{i+2}) \, ...
= \\ =  &... \, Q_{ \alpha_{i-1} b}(\sigma_i) (R\cdot A^{[i+1]})_{b \alpha_{i+1}}(\sigma_{i+2}, \sigma_{i+1}) A_{\alpha_{i+1} \alpha_{i+2}}^{[i+2]}(\sigma_{i+2}) \, ... = \\ =  &... \, Q_{ \alpha_{i-1} b}(\sigma_i) (R\cdot A^{[i+1]})_{(b \sigma_{i+1}), (\alpha_{i+1} \sigma_{i+2}) } A_{\alpha_{i+1} \alpha_{i+2}}^{[i+2]}(\sigma_{i+2}) \, ... = \\
=  &... \, Q_{ \alpha_{i-1} b}(\sigma_i) Q'_{bc}(\sigma_{i+1}) R'_{c \alpha_{i+1}}(\sigma_{i+2}) A_{\alpha_{i+1} \alpha_{i+2}}^{[i+2]}(\sigma_{i+2}) \, ... = \\
=  &... \, Q_{ \alpha_{i-1} b}(\sigma_i) Q'_{bc}(\sigma_{i+1}) A'_{c \alpha_{i+2}}(\sigma_{i+2})  \, ... 
\end{split}
\end{align*}
where the new index $c$ runs from 1 to $2 \chi$. All these steps are represented in Figure \ref{fig:exampledecomposition}. Now, it becomes obvious that, by means of these tricks, one can easily re-write $\psi^{(1)}[A,F](\pmb{\sigma})$
as a sum of MPS with maximum bond dimension equal to $2\chi$. By summing these $N$ MPS, we will get an MPS with maximum bond dimension $2 N \chi$, which can be eventually compressed. With similar tricks, higher order terms can be formally recast in an MPS with locally larger bond dimension. In particular, it is not difficult to realize that $\psi^{(n)}$ contains terms, as the one represented in Figure \ref{fig:order_n}, that give rise to a local bond dimension $2^n \chi$. 

\section{MPBS encoding volume law for EE}
The aim of this section is to show, by means of an explicit example, that MPBS can efficiently encode a volume law for the entanglement-entropy (EE), thus representing an excellent generalization of MPS. 
In particular, we will focus on MPBS as readjusted to be applied on 2D systems. For this purpose, we will first consider an example of Restricted Boltzman Machine (RBM) state given in \cite{deng2017}, showing that it can be rewritten as MPBS. Then, we will follow the proof given in \cite{deng2017} to demonstrate that this wave function, with a particular choice of the parameters, can encode a volume law for EE in the two dimensional geometry. To begin, let us write the generic expression for the RBM representation of a quantum state $\psi$, i.e.
\begin{equation*}
\psi_{\text{RBM} }( \pmb{\sigma} )= \sum_{ \{ h \} }  \exp \bigg[ \sum_{i=1}^N a_i \sigma_i + \sum_{m=1}^M h_m b_m + \sum_{i,m} W_{im} \sigma_i h_m \bigg] \, \, ,
\end{equation*}
where $h_m= \pm 1$, $m=1,2... \, M$ are the so-called hidden-variables and $a,b,W$ are the parameters of the network. We will set $a_i=0$, $\forall i$. The parameters $W$ give rise to connections between visible spin variables and hidden spin variables. In the case in which the connections are short-range, an area law for the EE can be proven \cite{deng2017}. However, when the connections are long-range, this is no longer true. \\

\begin{figure}[t!]
\centering
\includegraphics[width=0.4\textwidth]{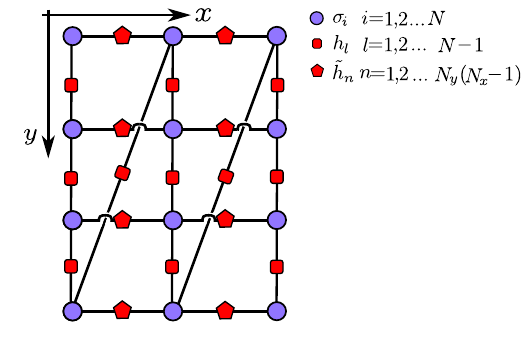}
\caption{
Graphical representation of the RBM state considered in the state. Visible layer variables (blue dots) are arranged in a two dimensional grid of shape $N_x, N_y$. Hidden variables (red shapes) give rise to connections. 
\label{fig:neurons}}
\end{figure}

For our purpose, let us set $N=N_x N_y$ and $M=(N \! - \! 1) \! + \! N_y ( N_x - 1 )=2N\!-\!N_y  - 1$. It is useful to split the hidden neurons in two sets: $h_l$ ($l=1,2... \,N \!- \! 1$) and $\tilde{h}_n$ ($n=1,2 ... \,  N_y ( N_x - 1 )$),
and define the respective connection parameters as
$
W_{il} = W(\delta_{i,l}+\delta_{i,l+1})
$
and
$
\tilde W_{in} = W(\delta_{i,n}+\delta_{i,n+N_y})
$.
The physical meaning under these choices is schematically represented in Figure \ref{fig:neurons}, where visible (hidden) variables are colored blue (red) and black lines represents $W$ connections. The idea is to use $\tilde{h}_n$ variables to give rise to MPBS terms connecting spins on the same row but different columns, whereas $h_l$ variables will give rise to the "snaking path" structure of the MPBS. \\ 

The wave function becomes the product of two terms ($\psi_{\text{RBM} }( \pmb{\sigma} )=\phi_1 ( \pmb{\sigma} ) \phi_2( \pmb{\sigma} ) $):
\begin{align*}
\begin{split}
\phi_1( \pmb{\sigma} ) =\prod_{l=1}^{N-1} \bigg( 2 \cosh \big[ b_l + W (\sigma_{l} + \sigma_{l+1} ) \big] & \bigg) = \prod_{l=1}^{N-1}  T_1 (\sigma_l, \sigma_{l+1}) \\
\phi_2 ( \pmb{\sigma} ) = \prod_{l=1}^{N_y ( N_x - 1 )}  \bigg( 2 \cosh \big[ \tilde{b}_l + W (\sigma_{l} + \sigma_{l+N_y} ) \big] &\bigg) = \prod_{l=1}^{N_y ( N_x - 1 )} T_2(\sigma_{l},\sigma_{l+N_y}) \, \, .\\
\end{split}
\end{align*}
 Let us consider the first one. We can set $T_1 (\sigma_l, \sigma_{l+1}) = \big(\pmb{v}^{(l)} \big)^T  \pmb{w}^{(l+1)}$, 
where 
\begin{align*}
\begin{split}
\big(\pmb{w}^{(l+1)}\big)^T &= \big( \exp \big[ b + W \sigma_{l+1}  \big] , \, \, \, \exp \big[ - b - W_{l} \sigma_{l+1}  \big] \big)  \\
\big(\pmb{v}^{(l)})^T &= \big( \exp \big[ W \sigma_{l}  \big] , \, \, \, \exp \big[ - W \sigma_{l}  \big] \big) \, \, .
\end{split}
\end{align*}
Thus, we get
\begin{align*}
\begin{split}
\phi_1( \pmb{\sigma} ) =  \big(\pmb{v}^{(1)} \big)^T  \pmb{w}^{(2)}  \big(\pmb{v}^{(2)} \big)^T  \pmb{w}^{(3)} ...   \big(\pmb{v}^{(N-1)} \big)^T  \pmb{w}^{(N)} = A^{[1]} A^{[2]} \, ... \, \, A^{[N]}  \, \, \, , \\
\end{split}
\end{align*}
where we defined the following matrices
\begin{equation*}
A^{[l]} =
\begin{cases}
\big(\pmb{v}^{(1)} \big)^T \, \, \text{ if } \, \, l=1 \\
\pmb{w}^{(l)} \big(\pmb{v}^{(l)}\big)^T \, \, \text{ if } \, \, l=2... \, N-1 \\
\pmb{w}^{(N)} \, \, \text{ if } \, \, l=N \\
\end{cases}
\end{equation*}
Let us observe that $A^{[l]}$ depends only on the local physical variable $\sigma_l$, therefore $\phi_1( \pmb{\sigma} )$ is in the form of an MPS, with bond dimension $\chi=2$. The MPS follows the ``snaking-path'', as displayed in Figure \ref{fig:neurons}. The wave function can now be written as
\begin{equation*}
\psi_{\text{RBM} }( \pmb{\sigma} ) = \pmb{A}(\sigma_1) T(\sigma_1, \sigma_{1+N_y}) \pmb{A}(\sigma_2) T(\sigma_2, \sigma_{1+N_y}) ... \, \pmb{A}(\sigma_{N-N_y}) T(\sigma_{N-N_y}, \sigma_{N}) \, \pmb{A}(\sigma_{N-N_y+1}) ...  \, \pmb{A}(\sigma_{N}) \, \, ,
\end{equation*}
where we used bold letters to distinguish the matrices (i.e.\ objects with two virtual indices) from the scalars (i.e.\ objects with no virtual indices).   
We can reabsorb the scalars in the matrices $\pmb{A}$, obtaining
\begin{equation}\label{eq:finex}
\psi_{\text{RBM} }( \pmb{\sigma} ) = \big( \pmb{A}(\sigma_1) +\pmb{F}^{[1]} (\sigma_1, \sigma_{1+N_y}) \big) \ ... \, \big( \pmb{A}(\sigma_{N-N_y}) +\pmb{F}^{[N-N_y]} (\sigma_{N-N_y}, \sigma_{N}) \big)  \pmb{A}(\sigma_{N-N_y+1}) ...  \, \pmb{A}(\sigma_{N}) \, \, , 
\end{equation}
where we defined the $\pmb{F}$ matrices as follows 
\begin{equation*}
\pmb{F}^{[l]}(\sigma_l, \sigma_{l+N_y}) = \pmb{A}(\sigma_l) \big( \sqrt{T(\sigma_l, \sigma_{l+N_y})} - 1 \big)  \, \, 
\end{equation*}
for $l=1,2... N-N_y$. Equation \ref{eq:finex} is just a particular case of our MPBS representation. Let us now set the RBM parameters as follows: $ b_l=- \frac{i \pi}{4}$, 
$\tilde{b}_n= \frac{i \pi}{2}$ and $W=\frac{i \pi}{4}$. It is not difficult to realize that with this choice, 
$\phi_1( \pmb{\sigma} ) = \pm c , \, \forall  \pmb{\sigma}$, and 
$
T_2(\sigma, \sigma') = \pm c' \delta_{\sigma,\sigma'}
$,
where $c$ and $c'$ are numerical constants and the sign depends on the spins. 
Thus, the RMB state will take the form 
\begin{align}\label{eq:rbmres}
\ket{ \psi_{\text{RBM} }}= \sum_{  \pmb{\sigma}_c } \pm C \ket{ \pmb{\sigma}_c ... \, \pmb{\sigma}_c } \, ,
\end{align}
with $C$ being a constant, and
where we used $\pmb{\sigma}_c$ to indicate the spin configuration of the first column (i.e.\ $\pmb{\sigma}_c=(\sigma_1, \sigma_2 ... \sigma_{N_y} )$). Equation \ref{eq:rbmres} means
that the state $\ket{ \psi_{\text{RBM} }}$ is the equal weights superposition of all the basis states corresponding to spin configurations in which the $N_x$ columns have all the same configuration $\pmb{\sigma}_c$. \\

\begin{figure}[t!]
\centering
\includegraphics[width=0.3\textwidth]{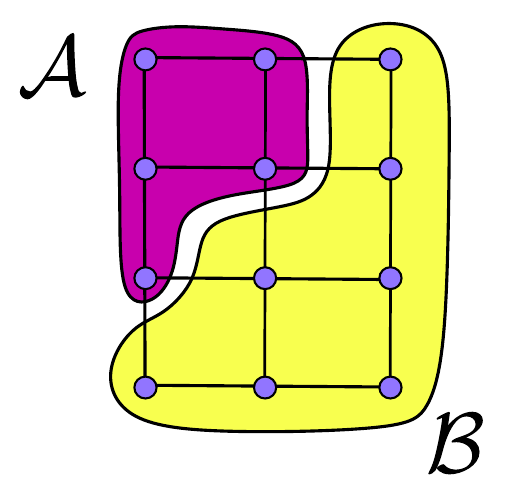}
\caption{Bipartition of a 2D system in two sub-systems $\mathcal{A}$ and $\mathcal{B}$. In the text, lattice sites are labelled by using the ''snaking path" convention.
\label{fig:split_system}}
\end{figure}

Let us now consider a bipartition of
our 2D system $\mathcal{A}$ and $\mathcal{B}$ (see Figure \ref{fig:split_system}). The subsystem density matrix $\rho_{\mathcal{A}}$ can be computed as
\begin{align*}
\begin{split}
\rho_{\mathcal{A}} = \text{tr}_{\mathcal{B}} [\ket{ \psi_{\text{RBM} }( \pmb{\sigma} ) } \bra{ \psi_{\text{RBM} }( \pmb{\sigma} ) } ] =
\sum_{  \pmb{\sigma}_c , \pmb{\sigma}_c'} C^2  \sum_{ \pmb{\sigma}_{\mathcal{B}}} \braket{ \pmb{\sigma}_{\mathcal{B}} |\pmb{\sigma}_c ...  \pmb{\sigma}_c } \braket{ \pmb{\sigma}_c' ...  \pmb{\sigma}_c' |  \pmb{\sigma}_{\mathcal{B}}} \, \, .
\end{split}
\end{align*}
If subsystem $\mathcal{B}$ contains at least one whole column, then fixing $\pmb{\sigma}_c$ also fix $\pmb{\sigma}_{\mathcal{B}}$ (and $\pmb{\sigma}_c'$). Therefore, we get
\begin{align*}
\begin{split}
\rho_{\mathcal{A}} \propto \sum_{  \pmb{\sigma}_c} \ket{\pmb{\sigma}_{\mathcal{A}} } \bra{\pmb{\sigma}_{\mathcal{A}} } \, .
\end{split}
\end{align*}
By reintroducing
the overall normalization constant, we get
$\rho_{\mathcal{A}} = \mathbb{1}/2^{|\mathcal{A}|}$, where $|\mathcal{A}|$ indicates the total number of lattice sites within the region $\mathcal{A}$. This result shows that $\mathcal{A}$ and $\mathcal{B}$ are maximally entangled and therefore the Rényi entropies associated with the bipartition are
\begin{equation*}
S_{\alpha}(\mathcal{A}|\mathcal{B})= |\mathcal{A}| \ln 2 \, \, \, \, \, \, \, \forall \alpha  \, \, .
\end{equation*}
These observations provide us a simple example of a wave function in the form of an MPBS efficiently encoding a volume law for the EE in the 2D geometry (the number of non-zero parameters $A,F$ scale polinomially with $N$). 

\section{DMRG technicalities} 

It is well known that to employ DMRG with 1D periodic systems can be challenging. 
In this section, we will show how we performed our simulations. 

\subsubsection{1D Ising model with Periodic Boundary Conditions}
Let us consider the 1D short-range Transverse Field Periodic Ising Chain (TFPIC). Let us notice that by defining the following matrices of operators 
\begin{equation}
W^{[1]} = 
    \begin{pmatrix}
    -h \sigma^x_1 & J \sigma^z_1 & \mathbb{1}_1 \\
    0 & 0 & -\sigma^z_1 \\
    0 & 0 & 0 \\    
    \end{pmatrix}
\, , \qquad W^{[i]} = 
    \begin{pmatrix}
    \mathbb{1}_i & 0 & 0 \\
    - \sigma^z_i & 0 & 0 \\
    -h \sigma^x_i & J \sigma^z_i & \mathbb{1}_i \\    
    \end{pmatrix}    \quad i=2,3 ... N
\end{equation}
we can rewrite the TFPIC hamiltonian in the form of a periodic MPO, i.e. \ 
\begin{equation}
    \text{tr} \bigg[W^{[1]} W^{[2]} ... \, W^{[N]} \bigg] = - J \sum_{i=1}^{N-1} \sigma^z_{i}  \sigma^z_{i+1} - J \sigma^z_{N}  \sigma^z_{1} - h \sum_{i=1}^N \sigma^x_i \, \, .
\end{equation}
The trace can be trivially re-written as a sum over three contractions
\begin{equation}\label{eq:contr}
    \text{tr} \bigg[W^{[1]}  ... \, W^{[N]} \bigg] = \big(v^{(1)}\big)^T W^{[1]}  ... \, W^{[N]} v^{(1)} + \big(v^{(2)}\big)^T W^{[1]}  ... \, W^{[N]} v^{(2)} + \big(v^{(3)}\big)^T W^{[1]}  ... \, W^{[N]} v^{(3)} \, \, ,
\end{equation}
where 
\begin{equation*}
    v^{(1)}= \begin{pmatrix}
    1 \\
    0 \\
    0 \\
    \end{pmatrix}
    \quad 
    v^{(2)}= \begin{pmatrix}
    0 \\
    1 \\
    0 \\
    \end{pmatrix}
   \quad 
    v^{(3)}= \begin{pmatrix}
    0 \\
    0 \\
    1 \\
    \end{pmatrix}
\end{equation*}
are the usual basis-vectors of $\mathbb{R}^3$. It is not difficult to realize that the third term of eq. \ref{eq:contr}, i.e.\ the contraction with $v^{(3)}$, vanishes. Thus, we obtain 
\begin{equation}\label{eq:mpob6}
    \text{tr} \bigg[W^{[1]} ... \, W^{[N]} \bigg] = \pmb{V}^{[1]} \pmb{W}^{[2]}  ... \, \pmb{W}^{[N-1]} \pmb{V}^{[N]} \, \,  \, ,
\end{equation}
where we defined the following tensors
\begin{equation*}
\pmb{W}^{[i]} = W^{[i]} \bigoplus W^{[i]} =\begin{pmatrix}
W^{[i]} & 0 \\
0 & W^{[i]} \\
\end{pmatrix} \, \, ,
\qquad \pmb{V}^{[N]} = \pmb{W}^{[N]} \begin{pmatrix}
v^{(1)} \\
v^{(2)} \\
\end{pmatrix} \, \, ,
\qquad \pmb{V}^{[1]} =  \begin{pmatrix}
\big(v^{(1)}\big)^T & \big(v^{(2)}\big)^T\\
\end{pmatrix} \pmb{W}^{[1]} \, \, . 
\end{equation*}
Equation \ref{eq:mpob6} is a representation of the TFPIC hamiltonian as an MPO with OBC and bond dimension equals to $6$. DMRG can now be applied in a straightforward way. Another possible trick is to write the TFPIC hamiltonian as an MPO of bond dimension $N+1$ and then apply an iterative compression alghoritm. However, our scheme seems to be much more simpler and fast.  \\

\subsubsection{Modified Haldane-Shastry model}

\begin{figure}
\centering
\includegraphics[width=0.5\textwidth]{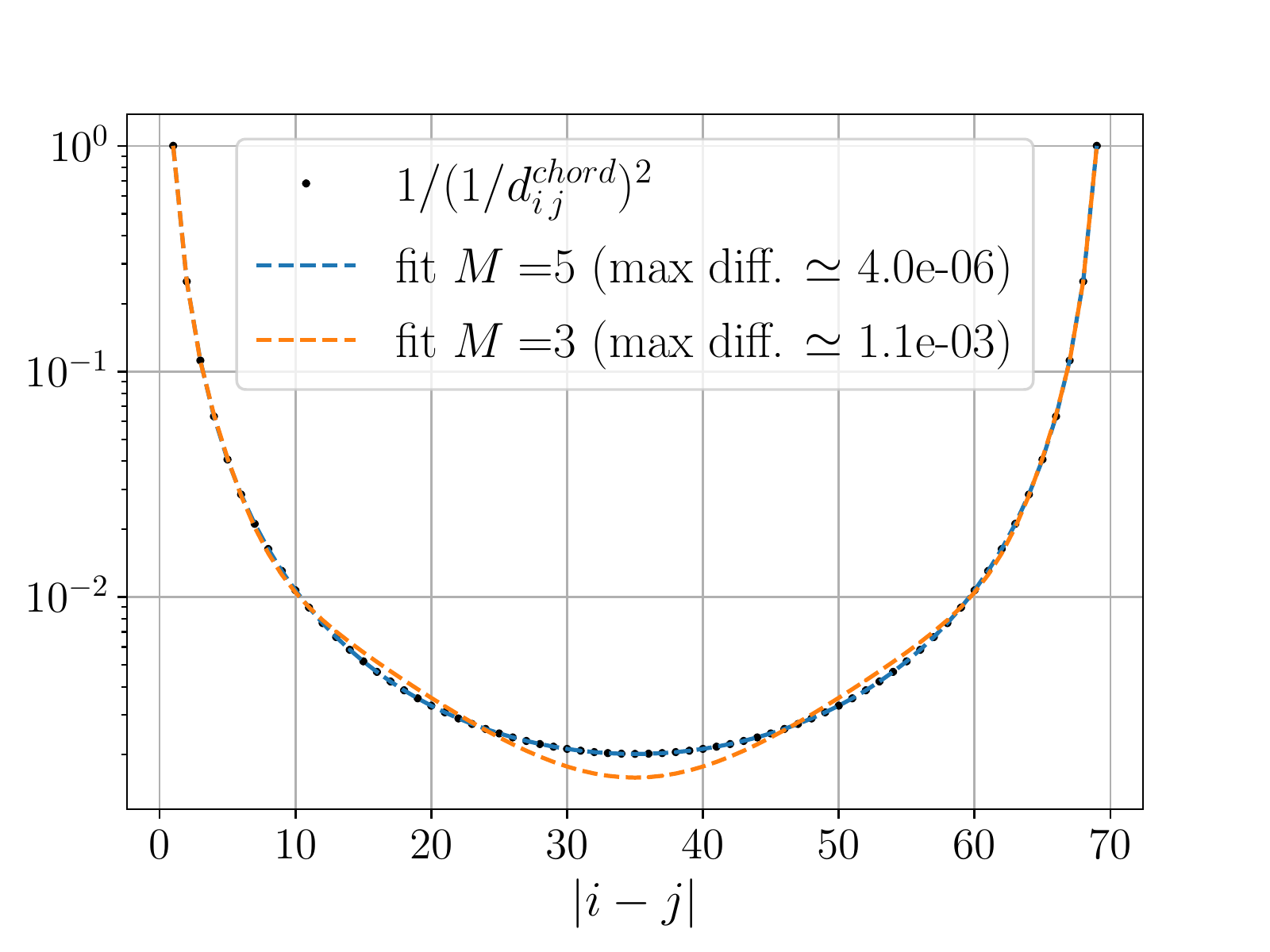}
\caption{Fit of the HS couplings $(1/d^{\text{chord}}_{ij})^2$ as a sum of $M$  symmetrically  decreasing/increasing exponentials. The values shown in the legend represent the maximum values of the differences between the fit and the HS couplings.    
\label{fig:HSfit}}
\end{figure}

Let us now consider the HS model. In this case, we fitted the couplings $(1/\widetilde{d}_{ij})^2$ with a sum of $M>1$ symmetrically decreasing and increasing exponentials, i.e.\ 
\begin{equation}
    \bigg(\frac{1}{\widetilde{d}_{ij}} \bigg)^2 \simeq \sum_{m=1}^M c_m \bigg(  e^{-|i-j|/\xi_m} +  e^{-N/\xi_m} e^{+|i-j|/\xi_m} \bigg) \, \, ,
\end{equation}
where $\xi_m$ are the characteristic lengths of the exponentials and $c_m$ are numerical coefficients. Let us observe that both the sides of the equality are symmetric under $|i\!-\!j| \! \leftrightarrow \! N-|i\!-\!j|$. The results of this fit are excellent, also for small values of $M$ (see Figure \ref{fig:HSfit}). By using these, we can rewrite the HS hamiltonian 
as
\begin{equation*}
    \mathcal{H}_{HS} \simeq \sum_{m=1}^M  \sum_{j < i} \big( c_m \lambda_m^{-|i-j|} + c_m \lambda_m^{-N} \lambda_m^{+|i-j|} \big)  \big( - \sigma^x_i \sigma^x_j - \sigma^y_i \sigma^y_j + \sigma^z_i \sigma^z_j \big) \, \, . 
\end{equation*}
It is well known how to rewrite this operator as an MPO \cite{schollwoeck2011}. In this case the bond dimension will be
$3M+2$.

\end{appendices}

\end{document}